\documentstyle[psfig,twoside,fleqn,espcrc2]{article}

\newcommand{\AmS}{{\protect\the\textfont2
  A\kern-.1667em\lower.5ex\hbox{M}\kern-.125emS}}
\def\gsim{\;
\raise0.3ex\hbox{$>$\kern-0.75em\raise-1.1ex\hbox{$\sim$}}\;
}
\def\lsim{\;
\raise0.3ex\hbox{$<$\kern-0.75em\raise-1.1ex\hbox{$\sim$}}\;
}
\newcommand {\ignore}[1]{}

\newcommand{\bc}{\begin{center}}
\newcommand{\ec}{\end{center}}

\def\ifmath#1{\relax\ifmmode #1\else $#1$\fi}

\def\3quarter{{\textstyle{3 \over 4}}}

\def\ra{\rightarrow}

\def\lf{\leaders\hbox to 1em{\hss.\hss}\hfill}
\def\21{$SU(2) \ot U(1)$}
\def\321{$SU(3) \ot SU(2) \ot U(1)$}
\def\ne{\hbox{$\nu_e$ }}
\def\nm{\hbox{$\nu_\mu$ }}
\def\nt{\hbox{$\nu_\tau$ }}
\def\ns{\hbox{$\nu_{s}$ }}

\def\O{\hbox{$\cal O$ }}

        \def\etc{\hbox{\it etc. }}

\def\neus{\hbox{neutrinos }}

\def\neu{\hbox{neutrino }}

\def\eq#1{{eq. (\ref{#1})}}

\def\VEV#1{\left\langle #1\right\rangle}

\def\ltap{\raisebox{-.4ex}{\rlap{$\sim$}} \raisebox{.4ex}{$<$}}

\def\lsim{\raise0.3ex\hbox{$\;<$\kern-0.75em\raise-1.1ex\hbox{$\sim\;$}}}
\def\gsim{\raise0.3ex\hbox{$\;>$\kern-0.75em\raise-1.1ex\hbox{$\sim\;$}}}

\def\beq{\begin{equation}}
\def\eeq{\end{equation}}
\def\bef{\begin{figure}}
\def\eef{\end{figure}}
\def\bet{\begin{table}}
\def\eet{\end{table}}
\def\bea{\begin{eqnarray}}
\def\ba{\begin{array}}
\def\ea{\end{array}}
\def\bi{\begin{itemize}}
\def\ei{\end{itemize}}
\def\ben{\begin{enumerate}}
\def\een{\end{enumerate}}
\def\ra{\rightarrow}
\def\ot{\otimes}
\def\eea{\end{eqnarray}}
\def\apj#1#2#3{          {\it Astrophys. J. }{\bf #1} (19#2) #3}

\def\ib#1#2#3{           {\it ibid. }{\bf #1} (19#2) #3}
\def\nat#1#2#3{          {\it Nature }{\bf #1} (19#2) #3}
\def\nps#1#2#3{        {\it Nucl. Phys. B (Proc. Suppl.) }{\bf #1} (19#2) #3} 
\def\np#1#2#3{           {\it Nucl. Phys. }{\bf #1} (19#2) #3}
\def\pl#1#2#3{           {\it Phys. Lett. }{\bf #1} (19#2) #3}
\def\pr#1#2#3{           {\it Phys. Rev. }{\bf #1} (19#2) #3}
\def\prep#1#2#3{         {\it Phys. Rep. }{\bf #1} (19#2) #3}
\def\prl#1#2#3{          {\it Phys. Rev. Lett. }{\bf #1} (19#2) #3}

\def\rmp#1#2#3{          {\it Rev. Mod. Phys. }{\bf #1} (19#2) #3}
\def\zp#1#2#3{           {\it Zeit. fur Physik }{\bf #1} (19#2) #3}

\def\n.c.#1#2#3{         {\it Nuovo Cim. }{\bf #1} (19#2) #3}
\def\r.n.c.#1#2#3{       {\it Riv. del Nuovo Cim. }{\bf #1} (19#2) #3}
\def\sjnp#1#2#3{         {\it Sov. J. Nucl. Phys. }{\bf #1} (19#2) #3}

\def\mpl#1#2#3{          {\it Mod. Phys. Lett. }{\bf #1} (19#2) #3}

\def\ppnp#1#2#3{           {\it Prog. Part. Nucl. Phys. }{\bf #1} (19#2) #3}

\def\tp{these proceedings}

\begin{document}
\begin{titlepage}
\begin{center}
\hfill hep-ph/9602369
\vskip 2cm
\large
{\bf Neutrino Properties}\\
\vskip 1cm
{\bf J. W. F. Valle}
\footnote{valle@flamenco.uv.es; URL http://neutrinos.uv.es}
\vskip 1cm
Instituto de F\'{\i}sica Corpuscular - C.S.I.C.\\
Departament de F\'{\i}sica Te\`orica, Universitat de 
Val\`encia\\46100 Burjassot, Val\`encia, Spain
\vskip 2cm
\end{center}

\begin{center}

{\bf Abstract}

\end{center}
\noindent
A brief sketch is made of the present observational status 
of neutrino properties, with emphasis on the hints from solar 
and atmospheric neutrinos, as well as cosmological data on the 
amplitude of primordial density fluctuations. Implications of 
neutrino mass in particle accelerators, astrophysics and 
cosmology are discussed.

\vfill
\noindent
Invited talk at TAUP95, Toledo, September 1995.
\end{titlepage}

\title{Neutrino Properties}
\author{J. W. F. Valle\address{Instituto de F\'{\i}sica Corpuscular 
- C.S.I.C.\\Departament de F\'{\i}sica Te\`orica, Universitat de 
Val\`encia\\46100 Burjassot, Val\`encia, Spain}
\thanks{Supported by DGICYT under Grant number PB92-0084.}}
\begin{abstract}
A brief sketch is made of the present observational status 
of neutrino properties, with emphasis on the hints from solar 
and atmospheric neutrinos, as well as cosmological data on the 
amplitude of primordial density fluctuations. Implications of 
neutrino mass in particle accelerators, astrophysics and 
cosmology are discussed.

\end{abstract}

\maketitle
\section{INTRODUCTION}
\vskip .2cm

It is beyond any doubt that, although very successful,
our present standard \21 model leaves open too many 
fundamental issues in particle physics to be an ultimate 
theory of Nature. One of the most fundamental ones refers 
to the masses and properties of neutrinos. Apart from being 
a theoretical puzzle, in the sense that there is no principle 
that dictates that neutrinos are massless, as postulated in 
the standard model, nonzero masses may in fact be required 
in order to account for the data on solar and atmospheric 
neutrinos, as well as the dark matter in the universe. 
The implications of detecting nonzero neutrino masses 
could be very far reaching for the understanding of 
fundamental issues in particle physics, astrophysics, 
as well as the large scale structure of our universe.

One interesting aspect of most extensions of the standard 
model where neutrino have non-vanishing masses is that they 
may affect the physics of the electroweak sector in a very 
remarkable way, which can be experimentally tested. Some of
the ways to probe the corresponding physics at accelerator 
as well as underground experiments will be described.

\subsection{Laboratory Limits }
\vskip .2cm

The most model-independent of the laboratory limits on \neu mass 
are those that follow purely from kinematics, given as \cite{PGG95}
\beq
\label{1}
m_{\nu_e} 	\lsim 5 \: \rm{eV}, \:\:\:\:\:
m_{\nu_\mu}	\lsim 250 \: \rm{keV}, \:\:\:\:\:
m_{\nu_\tau}	\lsim 23 \: \rm{MeV}
\eeq
The improved limit on the \ne mass from beta decays was recently 
given by Lobashev \cite{Erice}, while that on the \nt mass comes 
from the ALEPH experiment \cite{eps95} and may be substantially 
improved at a future tau-charm factory \cite{jj}. 

In addition, there are limits on neutrino masses that follow 
from the non-observation of neutrino oscillations \cite{granadaosc}. 
The 90\% confidence level (C.L.) exclusion contours of neutrino 
oscillation parameters in the 2-flavour approximation are given
in Fig. \ref{oscil}, taken from ref. \cite{jjc}. Improvements 
are expected from the ongoing CHORUS and NOMAD experiments at 
CERN, with a similar proposal at Fermilab \cite{chorus}.
There are also good prospects for substantial progress at future 
long baseline experiments using CERN and Fermilab neutrino beams 
aimed at the Gran Sasso and Soudan underground facilities,
respectively.
\bef
\vglue -1.5cm
\psfig{file=oscil.ps,height=11cm,width=7cm}
\vglue -3cm
\caption{Limits on oscillation parameters.}
\vglue -0.5cm
\label{oscil}
\eef

Another important limit follows from the non-observation of 
neutrino-less double beta decay - ${\beta \beta}_{0\nu}$ - i.e. 
the process by which an $(A,Z-2)$ nucleus decays to $(A,Z) + 2 \ e^-$. 
This process would arise from the virtual exchange of a Majorana 
neutrino from an ordinary double beta decay process. Unlike the 
latter, the neutrino-less process violates lepton number and
its existence would indicate the Majorana nature of neutrinos.
Because of the phase space advantage, this process is a very 
sensitive tool to probe into the nature of neutrinos.
In fact, as shown in ref. \cite{BOX}, a non-vanishing 
${\beta \beta}_{0\nu}$ decay rate requires \neus to be majorana 
particles, {\sl irrespective of which mechanism induces it}. 
This establishes a very deep connection which, in some special 
models, may be translated into a lower limit on the \neu masses.
The negative searches for ${\beta \beta}_{0\nu}$ in $^{76}$Ge
and other nuclei leads to a limit of about two eV \cite{Avignone} on
the weighted average \neu mass parameter $\VEV{m}$
\beq
\label{bb}
\VEV{m} \lsim 1 - 2 \ eV
\eeq
depending to some extent on the relevant nuclear matrix elements 
characterising this process \cite{haxtongranada}. Improved sensitivity 
is expected from the upcoming enriched germanium experiments. 
Although rather stringent, this limit in \eq{bb} may allow relatively 
large \neu masses, as there may be strong cancellations between the 
contributions of different neutrino types. This happens automatically
in the case of a Dirac \neu as a result of the lepton number symmetry 
\cite{QDN}. 

\subsection{The Cosmological Density Limit }
\vskip .2cm

In addition to laboratory limits, there is a cosmological bound that 
follows from avoiding the overabundance of relic neutrinos \cite{KT}
\beq
\label{RHO1}
m_{\nu_\tau} \lsim 92 \: \Omega_{\nu} h^2 \: eV\:,
\eeq
where $\Omega_{\nu} h^2 \leq 1$ and the sum runs over all isodoublet 
neutrino species with mass less than $O(1 \: MeV)$. Here 
$\Omega_{\nu}=\rho_{\nu}/\rho_c$, where $\rho_{\nu}$ is the neutrino
contribution to the total density and $\rho_c$ is the critical density.
The factor $h^2$ measures the uncertainty in the determination of the
present value of the Hubble parameter, $0.4 \leq h \leq 1$. 
The factor $\Omega_{\nu} h^2$ is known to be smaller than 1.

For the $\nu_{\mu}$ and $\nu_{\tau}$ this bound is much more 
stringent than the corresponding laboratory limits \eq{1}. 

Recently there has been a lot of work on the possibility of
an MeV tau neutrino \cite{ma1,SF}. Such range seems to be an 
interesting one from the point of view of structure formation 
\cite{ma1,SF}. Moreover, it is theoretically viable as the
constraint in \eq{RHO1} holds only if \neus are stable on the 
relevant cosmological time scales. In models with spontaneous 
violation of total lepton number \cite{CMP} there are new 
interactions of neutrinos with the majorons which may cause
neutrinos to decay into a lighter \neu plus a majoron, 
for example \cite{fae},
\beq
\label{NUJ}
\nu_\tau \ra \nu_\mu + J \:\: .
\eeq
or have sizeable annihilations to these majorons,
\beq
\label{JJ}
\nu_\tau + \nu_\tau \ra J + J \:\: .
\eeq
The possible existence of fast decay and/or annihilation channels 
could eliminate relic neutrinos and therefore allow them to be heavier
than \eq{RHO1}. The cosmological density constraint on neutrino decay 
lifetime (for neutrinos lighter than 1 MeV or so) may be written as
\beq
\tau \ltap 1.5 \times10^7 (KeV/m_{\nu_\tau})^{2} yr \:,
\label{RHO2}
\eeq
and follows from demanding an adequate red-shift of the heavy neutrino
decay products. For neutrinos heavier than $\sim 1 \: MeV$, such as
possible for the case of $\nu_{\tau}$, the cosmological limit on the
lifetime is less stringent than given in \eq{RHO2}.
 
As we already mentioned the possible existence of non-standard 
interactions of neutrinos due to their couplings to the Majoron
brings in the possibility of fast invisible neutrino decays with 
Majoron emission \cite{fae}. These 2-body decays can be
much faster than the visible decays, such as radiative decays of 
the type $\nu' \ra \nu + \gamma$. As a result the Majoron decays
are almost unconstrained by astrophysics and cosmology. For
a more detailed discussion see ref. \cite{KT}. 

A general method to determine the Majoron emission decay rates 
of neutrinos was first given in ref. \cite{774}. The resulting 
decay rates are rather subtle \cite{774} and model dependent
and will not be discussed here. The reader may consult ref.
\cite{V,fae}. The conclusion is that there are many ways
to make neutrinos sufficiently short-lived that all mass values
consistent with laboratory experiments are cosmologically acceptable.
For neutrino decay lifetime estimates see ref. \cite{fae,V,RPMSW}.

\subsection{The Nucleosynthesis Limit}
\vskip .2cm

There are stronger limits on neutrino lifetimes and/or
annihilation cross sections arising from cosmological nucleosynthesis
considerations. If massive $\nu_\tau$'s are stable during nucleosynthesis 
($\nu_\tau$ lifetime longer than $\sim 100$ sec), one can constrain
their contribution to the total energy density from the observed 
amount of primordial helium. This bound can be expressed through 
an effective number of massless neutrino species ($N_\nu$). Using
$N_\nu < 3.4-3.6$, the following range of $\nu_\tau$ mass has been 
ruled out \cite{KTCS91,DI93}
\begin{equation}
0.5~MeV < m_{\nu_\tau} < 35~MeV
\label{cons1}
\end{equation}
If the nucleosynthesis limit is taken less stringent the limit
loosens somewhat. However it has recently been argued that 
non-equilibrium effects from the light neutrinos arising from
the annihilations of the heavy \nt's make the constraint stronger
and forbids all $\nu_\tau$ masses on the few MeV range. 

One can show that if the \nt is unstable during nucleosynthesis 
\cite{unstable} the bound on its mass is substantially weakened 
translated as a function of the assumed lifetime \cite{unstable}.

Even more important is the effect of neutrino annihilations \cite{DPRV}.
Fig. 2 gives the effective number of massless neutrinos equivalent 
to the contribution of massive neutrinos with different values of 
the coupling $g$ between $\nu_\tau$'s and $J$'s, expressed in units 
of $10^{-5}$. For comparison, the dashed line corresponds to the 
standard model $g=0$ case. One sees that for a fixed $N_\nu^{max}$, 
a wide range of tau neutrino masses is allowed for large enough 
values of $g$. No \nt masses below 23 MeV can be ruled out, as long 
as $g$ exceeds a few times $10^{-4}$. Such values are reasonable in many 
majoron models \cite{fae,MASIpot3}. For more details see ref. \cite{DPRV}.
\begin{figure}
\label{neq}
\psfig{file=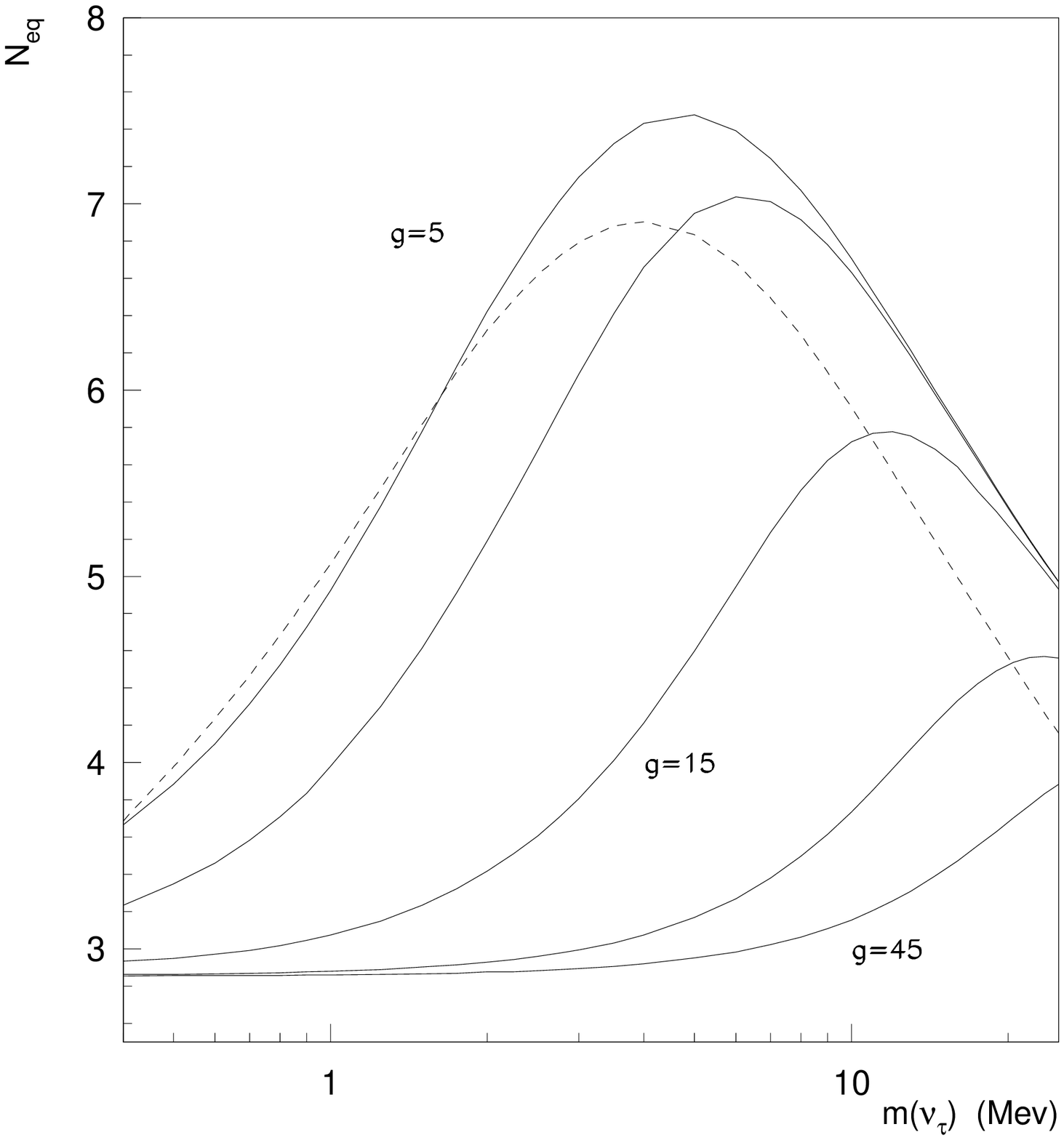,height=6cm,width=7cm}
\vglue -.5cm
\caption{Contribution of a heavy \nt to nucleosynthesis
in terms of the equivalent number of massless neutrinos.}
\end{figure}
In short one sees that the constraints on the mass of a Majorana
$\nu_\tau$ from primordial nucleosynthesis can be substantially 
relaxed if annihilations $\nu_\tau \bar{\nu}_\tau \leftrightarrow JJ$ 
are present. More details in the talk by Pastor \cite{Pastor}.

As a result of the above considerations one concludes that
it is worthwhile to continue the efforts to improve present 
laboratory \neu mass limits in the laboratory. One method
sensitive to large masses is to search for distortions in 
the energy spectra of leptons coming from $\pi, K$
weak decays such as $\pi, K \ra e \nu$, $\pi, K \ra \mu \nu$, 
as well as kinks in nuclear $\beta$ decays.

\section{HINTS FOR NEUTRINO MASSES}
\vskip .2cm

So far the only indications in favour of nonzero neutrino rest
masses have been provided by astrophysical and cosmological 
observations, with a varying degree of theoretical input.  
We now turn to these.

\subsection{Dark Matter}
\vskip .2cm

By combining the observations of cosmic background temperature 
anisotropies on large scales performed by the COBE  satellite 
\cite{cobe} with cluster-cluster correlation data e.g. from 
IRAS \cite{iras} one finds that it is not possible to fit
well the data on all scales within the framework of the
popular cold dark matter (CDM) model. Indeed, the best fit is 
obtained for a mixture, otherwise ad hoc, consisting of
about 70\% CDM with about 25 \% {\sl hot dark matter} (HDM)
and a small amount in baryons \cite{cobe2}.
The best way to make up for the hot dark matter component 
is through a massive neutrino in the few eV mass range.
It has been argued that this could be the tau neutrino,
in which case one might expect the existence of \ne $\ra$ \nt 
or \nm $\ra$ \nt oscillations. Searches for these oscillations
are now underway at CERN, with a similar proposal also at 
Fermilab \cite{chorus}. This mass scale is also consistent 
with the hints in favour of neutrino oscillations reported 
by the LSND experiment \cite{Caldwell}.

\subsection{Solar Neutrinos}
\vskip .2cm

So far the averaged data collected by the chlorine \cite{cl}, 
Kamiokande \cite{k}, as well as by the low-energy data on pp 
neutrinos from the GALLEX and SAGE experiments \cite{ga,sa} 
still pose a persisting puzzle. The most recent data can be
summarised as:
\bea
\label{data}
R_{Cl}^{exp}= (2.55 \pm 0.25) \mbox{SNU} \\ \nonumber
R_{Ga}^{exp}= (74 \pm 8) \mbox{SNU}  \\ \nonumber
R_{Ka}^{exp}= (0.44 \pm 0.06) R_{Ka}^{BP95} 
\eea 
where  $R_{Ka}^{BP95}$ is the BP95 SSM prediction of ref. \cite{SSM}.
For the gallium result we have taken the average of the GALLEX \cite{ga} 
and the SAGE measurements \cite{sa}. 

Comparing the data of gallium experiments with the Kamiokande data
one sees the need for a reduction of the $^7$Be flux relative to 
standard solar model \cite{SSM} expectations. Inclusion of the 
Homestake data only sharpens the discrepancy, suggesting that the 
solar \neu problem is indeed a real problem. The totality of the 
data strongly suggests that the simplest astrophysical solutions 
are ruled out, and that new physics is needed \cite{CF}. The most 
attractive possibility is to assume the existence of \neu 
conversions involving very small \neu masses. In the framework 
of the MSW effect \cite{MSW} the required solar neutrino 
parameters $\Delta m^2$ and $\sin^2 2\theta$ are determined 
through a $\chi^2$ fit of the experimental data
\footnote{For simplicity we neglect theoretical uncertainties,
earth effects, as well as details of the neutrino production 
region.}. 
Fig. 3, taken from ref. \cite{noise}, shows the 90\% C.L.
areas for the in the BP95 model for the case of active neutrino 
conversions. The fit favours the small mixing solution over the 
large mixing one, due mostly to the larger reduction of the 
$^7$Be flux found in the former.
\bef
\psfig{file=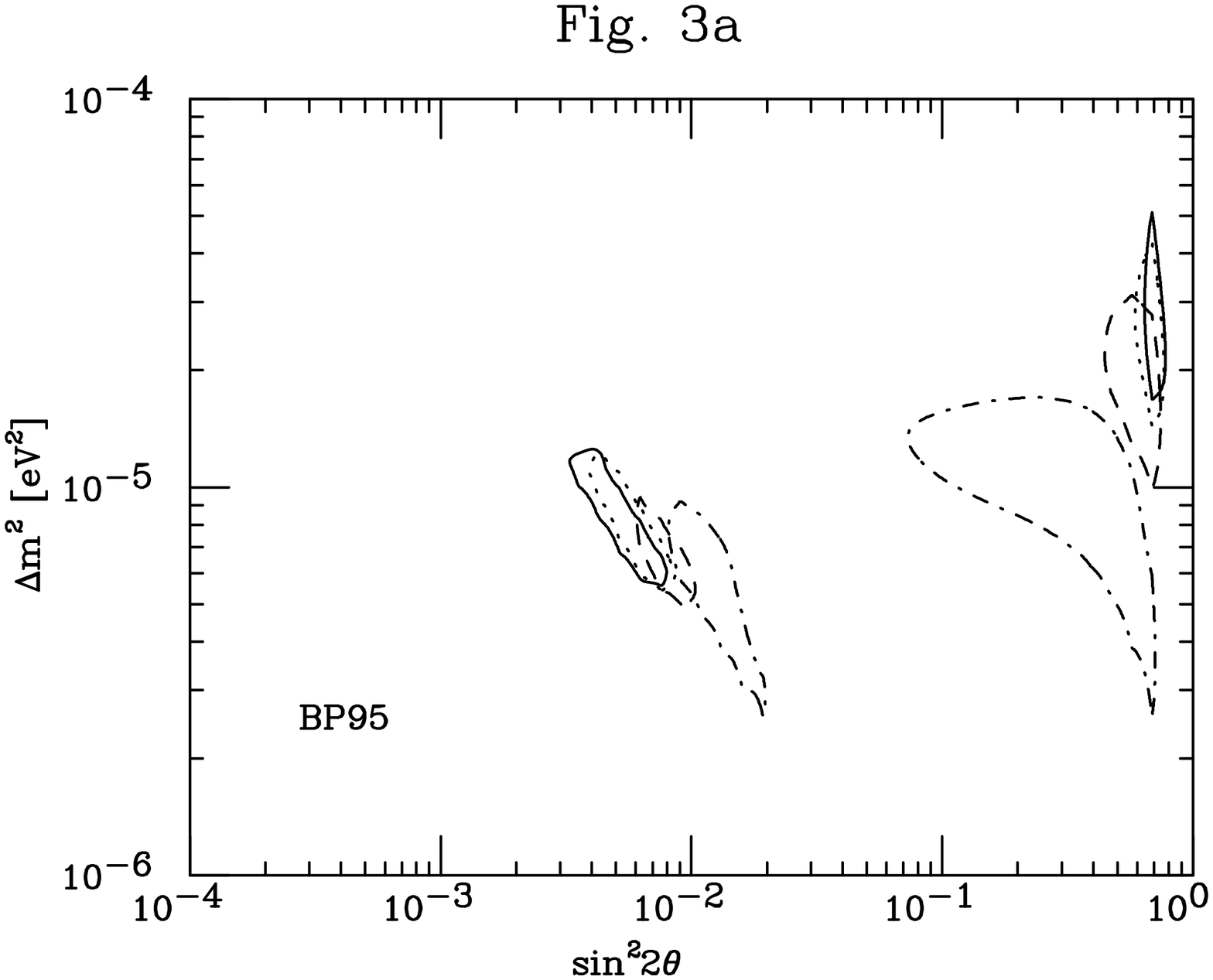,width=7.5cm,height=8cm,angle=90}
\vglue -1.3cm
\caption{Allowed solar neutrino oscillation parameters for
active neutrino conversions.}
\vglue -0.5cm
\label{msw}
\eef
Here $\xi$ denotes the assumed level of noise fluctuations in 
the solar matter density \cite{BalantekinLoreti}, not excluded by 
the SSM nor by present helioseismology studies. The solid curves 
are for the standard $\xi=0$ assumption corresponding to a smooth Sun. 
The regions inside the other curves correspond to the case
where matter density fluctuations are assumed. Noise causes
a slight shift of $\Delta m^2$ towards lower values and a 
larger shift of $\sin ^2 2 \theta$ towards larger values. 
The corresponding allowed $\Delta m^2$ range is 
$2.5 \times 10^{-6} <\Delta m^2< 9 \times 10^{-6}$ eV$^2$
instead of 
$5 \times 10^{-6} <\Delta m^2< 1.2 \times 10^{-5}$ eV$^2$
in the noiseless case.
The large mixing area is less stable, with a tendency to shift 
towards smaller $\Delta m^2$ and $\sin^2 2 \theta$ values.

It is interesting to note that the $^7$Be neutrinos are the 
solar neutrino spectrum component which is most affected by 
the matter noise. Therefore the Borexino experiment should 
be an ideal tool for studying the solar matter fluctuations, 
if sufficiently small errors can be achieved. Its potential in 
"testing" the level of solar matter density fluctuations is 
discussed in ref. \cite{noise}, summarized in the talk by Rossi 
\cite{rossi}. Ref. \cite{noise} also contains a discussion 
of sterile solar neutrino conversions, as well as a
comparison with other solar models.

\vglue -.8cm
\subsection{Atmospheric Neutrinos}
\vskip .2cm

Two underground experiments, Kamiokande and IMB, and possibly 
also Soudan2, have indications which support an apparent deficit 
in the expected flux of atmospheric $\nu_\mu$'s relative to that 
of $\nu_e$'s that would be produced from conventional decays of 
$\pi$'s, $K$'s as well as secondary muon decays \cite{Barish}. 
Although the predicted absolute fluxes of \neus produced by 
cosmic-ray interactions in the atmosphere are uncertain at the 
20\% level, their ratios are expected to be accurate to within 
5\%. While some of the experiments, such as Frejus and NUSEX, 
have not found a firm evidence, it has been argued that there 
may be a strong hint for an atmospheric neutrino deficit that 
could be ascribed to \neu oscillations.
Recent results from Kamiokande on higher energy \neus 
strengthen the case for an atmospheric \neu problem. The
relevant oscillation parameters are shown in Fig. \ref{kamglasgow}
taken from ref. \cite{atm}.
\bef
\psfig{file=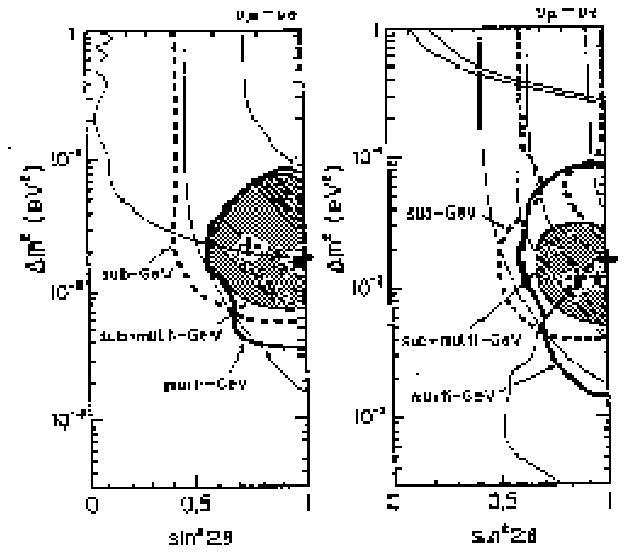,width=7.5cm,height=8cm}
\vglue -2cm
\caption{Atmospheric \neu oscillation parameters
from Kamiokande data.}
\vglue -0.5cm
\label{kamglasgow}
\eef

\vglue 1cm
\subsection{Reconciling Present Hints.}
\vskip .2cm

One of the simplest extensions of the electroweak theory 
consists in adding isosinglet neutral heavy leptons (NHLS), 
such as right handed neutrinos, as in the seesaw model \cite{GRS}. 
In this case the NHLS have a large Majorana mass term $M_R$, which 
violates total lepton number, or B-L (baryon minus lepton number), a 
symmetry that plays an important role in many extended gauge 
models \cite{LR}. The masses of the light neutrinos are
obtained by diagonalizing the following mass matrix
\beq
\begin{array}{c|cccccccc}
& \nu & \nu^c\\
\hline
\nu   & 0 & D \\
\nu^c & D^T & M_R
\end{array}
\label{SS}
\eeq
where $D = h_D v_2 /\sqrt2$ is the Dirac mass matrix and
$M_R = M_R^T$ is the isosinglet Majorana mass. In the seesaw 
approximation, one finds 
\beq
{M_L} = - D M_R^{-1} D^T \:.
\label{SEESAW}
\eeq
Although one expects $M_R$ to be large, one can not make any 
firm guess, as its magnitude heavily depends on the model. 
As a result one can not make any real prediction for the 
corresponding light neutrino masses that are generated 
through the exchange of the heavy Majorana neutrinos. 

\subsubsection{Almost Degenerate Neutrinos}
\vskip .2cm

The above observations from cosmology and astrophysics do 
seem to suggest a theoretical puzzle. As can easily be
understood just on the baisis of numerology, it seems rather 
difficult to reconcile the three observations discussed above 
in a framework containing just the three known \neus. The only 
possibility to fit these observations in a world with just the 
three known neutrinos is if all of them have nearly the same 
mass $\sim$ 2 eV \cite{caldwell}. This can be arranged, for
example in general seesaw models which also contain an effective 
triplet vacuum expectation value \cite{2227,LR} contributing to
the light neutrino masses. This term should be added to \eq{SEESAW}.
Thus one can construct extended seesaw models where the main 
contribution to the light \neu masses ($\sim$ 2 eV) is universal,
due to a suitable horizontal symmetry, while the splittings 
between \ne and \nm explain the solar \neu deficit and that 
between \nm and \nt explain the atmospheric \neu anomaly \cite{DEG}.

\subsubsection{Four Neutrino Models}
\vskip .2cm

The alternative way to fit all the data is to add a 
fourth \neu species which, from the LEP data on the 
invisible Z width, we know must be of the sterile type,
call it \ns. The first scheme of this type gives mass
to only one of the three neutrinos at the tree level,
keeping the other two massless \cite{OLD}. 
In a seesaw scheme with broken lepton number, radiative 
corrections involving gauge boson exchanges will give 
small masses to the other two neutrinos \ne and \nm
\cite{Choudhury}. However, since the singlet \neu is 
super-heavy in this case, there is no room to account 
for the three hints discussed above.

Two basic schemes have been suggested to keep the sterile
neutrino light due to a special symmetry. In addition to the
sterile \neu \ns, they invoke additional Higgs bosons beyond 
that of the standard model, in order to generate radiatively
the scales required for the solar and atmospheric \neu
conversions. In these models the \ns either lies at the dark matter 
scale \cite{DARK92} or, alternatively, at the solar \neu scale 
\cite{DARK92B}. 
In the first case the atmospheric
\neu puzzle is explained by \nm to \ns oscillations,
while in the second it is explained by \nm to \nt
oscillations. Correspondingly, the deficit of
solar \neus is explained in the first case
by \ne to \nt oscillations, while in the second 
it is explained by \ne to \ns oscillations. In both 
cases it is possible to fit all observations together. 
However, in the first case there is a clash with the 
bounds from big-bang nucleosynthesis. In the latter 
case the \ns is at the MSW scale so that nucleosynthesis
limits are satisfied. They nicely agree with the best 
fit points in Fig. \ref{kamglasgow}, taken from 
ref. \cite{atm}. Moreover, it can naturally fit the 
recent preliminary hints of neutrino oscillations of 
the LSND experiment \cite{Caldwell}.

Another theoretical possibility is that all active
\neus are very light, while the sterile \neu \ns is
the single \neu responsible for the dark matter
\cite{DARK92D}.
\vglue -1cm

\subsubsection{Mev Tau Neutrino}
\vskip .2cm

An MeV range tau neutrino is an interesting possibility 
to consider for two reasons. First, such mass is within 
the range of the detectability, for example at a tau-charm
factory \cite{jj}. On the other hand, if such neutrino 
decays  before the matter dominance epoch, its decay
products would add energy to the radiation, thereby 
delaying the time at which the matter and radiation 
contributions to the energy density of the universe 
become equal. Such delay would allow one to reduce
the density fluctuations on the smaller scales purely 
within the standard cold dark matter scenario, and could 
thus reconcile the large scale fluctuations observed by
COBE \cite{cobe} with the observations such as those 
of IRAS \cite{iras} on the fluctuations on smaller scales.

In ref. \cite{JV95} a model was presented where an unstable 
MeV Majorana tau \neu naturally reconciles the cosmological 
observations of large and small-scale density fluctuations 
with the cold dark matter model (CDM) and, simultaneously, 
with the data on solar and atmospheric neutrinos discussed
above. The solar \neu deficit is explained through long 
wavelength, so-called {\sl just-so} oscillations involving 
conversions of \ne into both \nm and a sterile species \ns, 
while the atmospheric \neu data are explained through \nm 
$\ra$ \ne conversions. Future long baseline \neu oscillation 
experiments, as well as some reactor experiments will test 
this hypothesis. The model assumes the spontaneous violation
of a global lepton number symmetry at the weak scale. 
The breaking of this symmetry generates the cosmologically 
required decay of the \nt with lifetime
$\tau_{\nu_\tau} \sim 10^2 - 10^4$ seconds, as well 
as the masses and oscillations of the three light 
\neus \ne, \nm and \ns required in order to account for 
the solar and atmospheric \neu data. One can verify 
that the big-bang nucleosynthesis constraints 
\cite{KTCS91,DI93} can be satisfied in this model. 

\section{IMPLICATIONS}
\label{impli}
\vskip .2cm

There is a variety of new phenomena that could be associated 
with neutrino mass \cite{fae}. Although in the simplest models 
of seesaw type the NHLS are Majorana type and expected to 
be quite heavy, due to limits on the light \neu masses, there 
are interesting variants with light Dirac NHLS \cite{SST}.
After mass matrix diagonalization there are couplings connecting 
light to heavy neutrinos \cite{2227}, restricted only by present 
constraints on weak universality violation. Through these the NHLS 
can be singly produced in $Z$ decays, if their mass is below that 
of the $Z$ \cite{CERN}
\begin{equation}
Z \rightarrow N_{\tau} + \nu_{\tau}
\end{equation}
Subsequent NHL decays would then give rise to large missing 
momentum events, called zen-events. The attainable rates for 
such processes can lie well within the sensitivities of the 
LEP experiments \cite{CERN}. Dedicated searches for acoplanar 
jets and lepton pairs from $Z$ decays have provided stringent
constraints on NHL couplings to the $Z$, plotted below \cite{lacasta}
\footnote{There have been also inconclusive hints reported by 
ALEPH \cite{alephmono}.}
\bef
\psfig{file=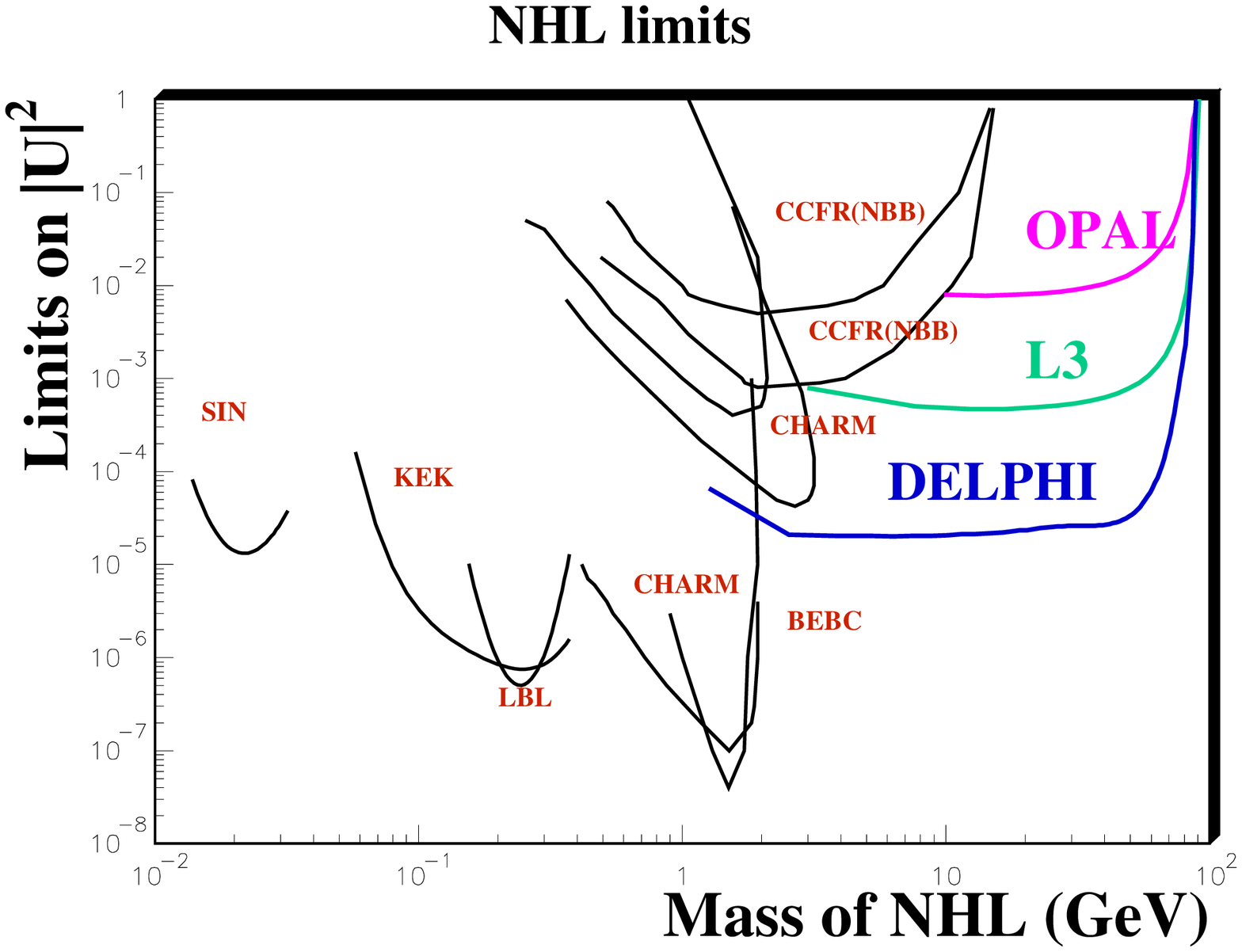,width=7.0cm,height=8cm}
\vglue -3cm
\caption{Limits on NHL mass and couplings.}
\label{nhlimits}
\eef 
One sees that the recent DELPHI constraints supersede by
far the low energy constraints following, e.g. from weak 
universality.

Even when the isosinglet neutral heavy leptons are heavier than 
the $Z$, they can produce interesting effects. For example, these 
NHLS may mediate lepton flavour violating (LFV) decays which are 
exactly forbidden in the standard model. These virtual effects 
are completely calculable in these models, in terms of the NHL 
masses and electroweak charged and neutral current couplings.
In the simplest models of seesaw type where the NHLS are 
Majorana type these decays are expected to be small, due to 
limits on the light \neu masses. However, in other variant
models with Dirac NHLS \cite{SST} this suppression is not 
present \cite{BER,CP} and LFV rates are restricted only
by present constraints on weak universality violation.
These allow for sizeable decay branching ratios, close
to present experimental limits \cite{3E} and within the 
sensitivities of the planned tau and B factories \cite{TTTAU}. 
The situation is summarised in the Tables.

\begin{table}
\begin{center}
\begin{displaymath}
\begin{array}{|c|cr|} 
\hline
\mbox{channel} & \mbox{strength} & \mbox{} \\
\hline
\tau \rightarrow e \gamma ,\mu \gamma &  \lsim 10^{-6} & \\
\tau \rightarrow e \pi^0 ,\mu \pi^0 &  \lsim 10^{-6} & \\
\tau \rightarrow e \eta^0 ,\mu \eta^0 &  \lsim 10^{-6} - 10^{-7} & \\
\tau \rightarrow 3e , 3 \mu , \mu \mu e, \etc &  \lsim 10^{-6} - 10^{-7} & \\
\hline
\end{array}
\end{displaymath}
\vglue 0.2cm
\caption{Allowed LFV $\tau$ decay branching ratios.}
\end{center}
\vglue -0.5cm 
\end{table}
The study of these rare $Z$ decays nicely complements what can 
be learned from the study of rare LFV muon and tau decays. 
The stringent limits on $\mu \rightarrow e \gamma$ preclude 
the corresponding process $Z \ra e \mu$ of being sizeable.
However the decays $Z \ra e \tau$ and $Z \ra \mu \tau$ can 
occur at the \O($10^{-6}$) level. Similar statements can be 
made also for the CP violating Z decay asymmetries in these 
LFV processes \cite{CP}. However, under realistic assumptions, 
it is unlikely that one will be able to see these decays at LEP 
without a high luminosity option \cite{ETAU}. In any case there 
have been dedicated searches which have set good limits in the 
range from $10^{-6}$ to $10^{-4}$ for LFV Z and $\tau$ decays 
at LEP \cite{opallfv}. 
\begin{table}
\begin{center}
\begin{displaymath}
\begin{array}{|c|cr|} 
\hline
\mbox{channel} & \mbox{strength} & \mbox{} \\
\hline
Z \rightarrow e \tau &  \lsim 10^{-6} - 10^{-7} & \\
Z \rightarrow \mu \tau &  \lsim 10^{-7} & \\
\hline
\end{array}
\end{displaymath}
\vglue 0.2cm
\caption{Allowed branching ratios for LFV $Z$ decays.}
\end{center}
\vglue -0.5cm 
\end{table}
Finally we note that there can also be large rates for lepton 
flavour violating decays in models with radiative mass generation 
\cite{zee.Babu88}. The expected decay rates may lie within the 
present experimental sensitivities and the situation should 
improve at PSI or at the proposed tau-charm factories.

Some models of massive neutrinos may lead to quite important
and unexpected effects in the electroweak breaking sector
which may contain a massless Nambu-Goldstone boson, denoted
by $J$, in the physical spectrum \cite{CMP,fae}. This leads
to a new possibility that the Higgs bosons decay invisibly 
as \cite{JoshipuraValle92}
\beq
h \ra J + J
\eeq
The simplest mode of production of the Higgs boson at LEP
is through the Bjorken mechanism. The production and subsequent 
decay of a Higgs particle which may decay visibly or invisibly 
involves three independent parameters: its mass $M_H$, its 
coupling strength to the Z, normalized by that of the standard 
model, $\epsilon_B^2$, and its invisible decay branching ratio. 
One can use the LEP searches for various exotic channels 
in order to determine the regions in parameter space 
that are already ruled out \cite{alfonso}. 

The invisible decay of the Higgs boson may also affect the 
strategies for searches at higher energies. For example, the 
ranges of parameters that can be covered by LEP200 searches 
for various integrated luminosities and centre-of-mass energies 
have been investigated \cite{ebolepp2}, and the results are 
illustrated in Fig. \ref{bjlep2}.
\bef
\psfig{file=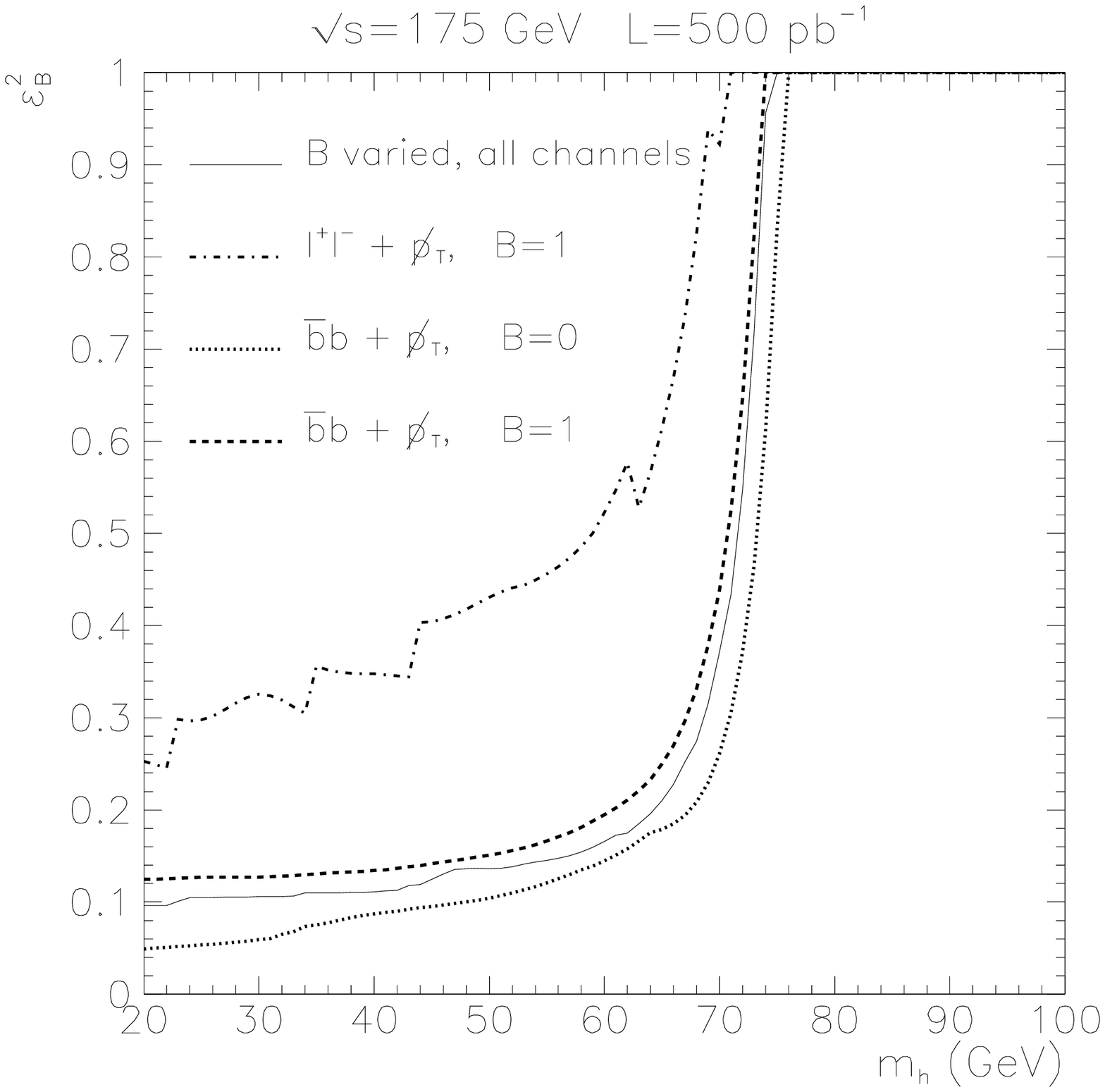,width=7.5cm,height=7cm}
\caption{Higgs mass and coupling that can be explored at LEP200.}
\vglue -0.5cm
\label{bjlep2}
\eef
Another mode of production of invisibly decaying Higgs bosons 
is that in which a CP even Higgs boson is produced in association 
with a massive CP odd scalar \cite{HA}. This production mode is 
present in all but the simplest majoron model containing just one 
complex scalar singlet in addition to the standard Higgs doublet. 
Present limits on the corresponding coupling strength parameter 
are given in Fig. \ref{halep2} as a function of the A and H 
masses, for the case of a visibly decaying A boson and an
invisibly decaying H boson. This figure is taken from ref. 
\cite{HA} which contains extensive discussion of various 
integrated luminosities and centre-of-mass energy assumptions. 
\bef
\psfig{file=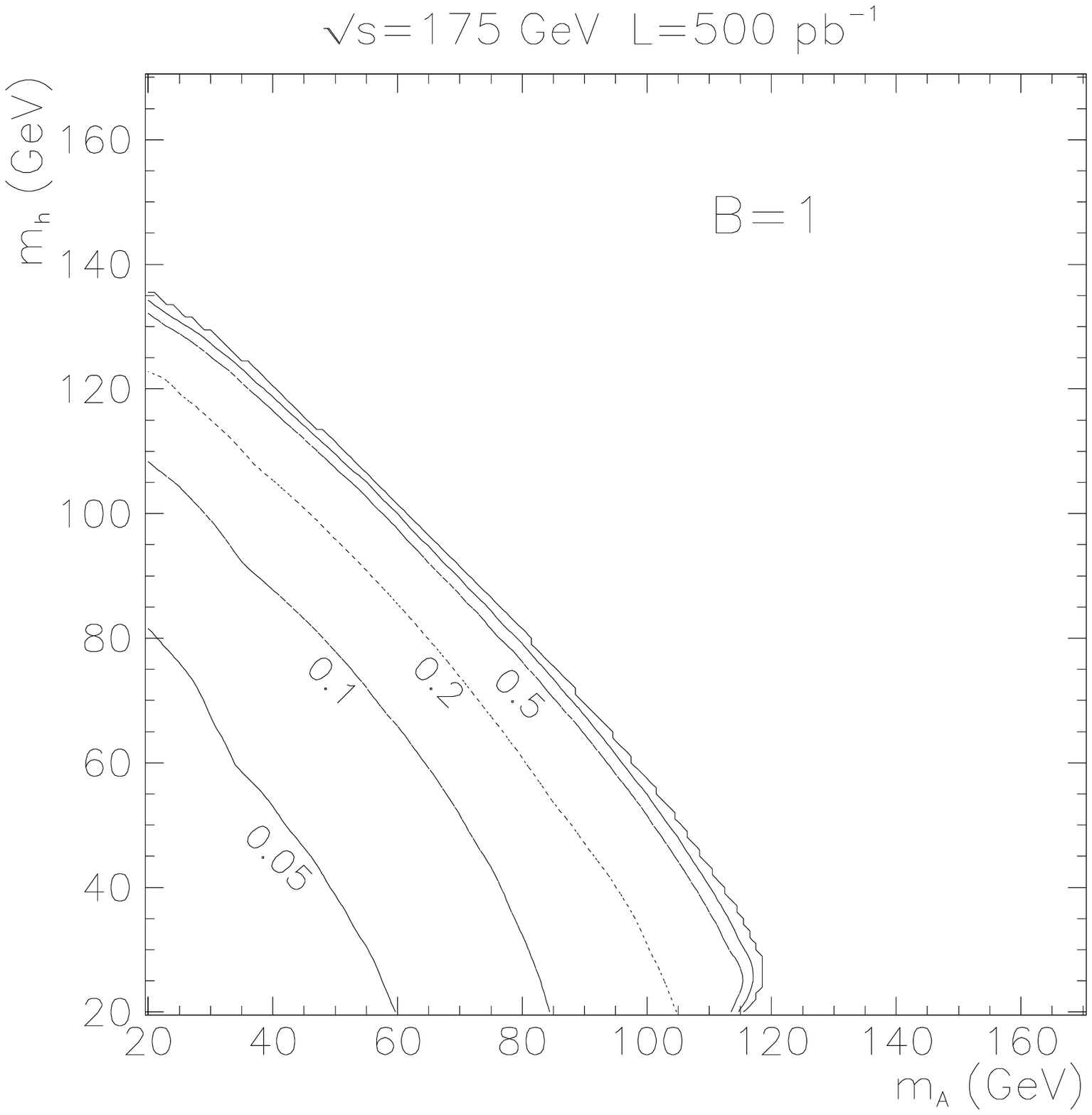,width=8cm,height=6.5cm}
\vglue -1cm
\caption{Higgs mass and coupling that can be explored at LEP200 
in the $e^+ e^- \ra H \:A$ production channel.}
\vglue -0.5cm
\label{halep2}
\eef
Similar analysis can be made for the case of a high energy linear 
$e^+ e^-$ collider (NLC) \cite{EE500}, as well as the LHC \cite{granada}.

\section{CONCLUSION}
\vskip .2cm

Theory can not yet predict fermion masses, and neutrinos are
no exception. Nevertheless neutrino masses are strongly suggested 
by present theoretical models of elementary particles. On the other 
hand, they seem to be required to fit together present astrophysical 
and cosmological observations. Neutrino mass studies in nuclear 
$\beta$ decays and peak search experiments should continue. 
Searches for $\beta \beta_{0\nu}$ decays with enriched germanium 
could test the quasi-degenerate neutrino scenario of section 2.4.1. 
Underground experiments at Superkamiokande, Borexino, and Sudbury 
will shed more light on the solar neutrino issue. 
Oscillation searches in the \ne $\ra$ \nt and \nm $\ra$ \nt 
channels at accelerators should soon improve over the present 
situation illustrated in Fig. 1, while long baseline experiments 
both at reactors and accelerators are being considered. These will 
test the regions of oscillation parameters presently suggested by 
atmospheric \neu data, shown in Fig. 4. Finally, new satellite 
experiments capable of measuring with better accuracy the cosmological 
temperature anisotropies at smaller angular scales than COBE, will
test different models of structure formation, and presumably shed 
light on the possible role of neutrinos as dark matter. 

If neutrinos are massive they could be responsible for a 
wide variety of implications, covering an impressive range 
of energies. These could be probed in experiments performed 
at underground installations as well as particle accelerators. 
For example, we saw in section 3 how neutrinos may produce
new signatures at high energy physics collider experiments. 
Although indirectly, these may test neutrino properties
in an important way and will therefore complement the efforts
at low energies as well as the non-accelerator studies.


\bibliographystyle{ansrt}

\begin{thebibliography}{99}

\bibitem{PGG95}
L. Montanet et al \pr{D50}{94}{1173} and 1995 off-year update for
the 1996 edition (URL: http://pdg.lbl.gov/).

\bibitem{Erice}
V. Lobashev, First Results of the Troitsk Experiment on the 
Search for Electron Neutrino Rest Mass in Tritium Decay,
preprint INR 862/94, 1994

\bibitem{eps95}
D. Buskulic et al., \pl{B349} {95} {585}.

\bibitem{jj}
J. G\'omez-Cadenas, M. C. Gonz\'alez-Garc\'{\i}a, \pr {D39} {89} {1370};
J. G\'omez-Cadenas et al., \pr{D41} {90} {2179}; see also
Third Workshop on the Charm Tau Factory, Marbella, Spain, June 1993,
(World Scientific, 1994), Ed. J. Kirkby and R. Kirkby.

\bibitem{granadaosc}
See e.g. J. Schneps, \nps{31}{93}{307}.

\bibitem{jjc}
J. G\'omez-Cadenas, M. C. Gonz\'alez-Garc\'{\i}a, hep-ph/9504246

\bibitem{chorus}
CHORUS and NOMAD proposals CERN-SPSLC/91-21 (1992) and CERN-SPSC/90-42 
(1992); J. G\'omez-Cadenas, J. A. Hernando, A. Bueno, CERN-PPE/95-177;
K. Kodama et~al., FNAL proposal P803 (1991)

\bibitem{BOX}
 J. Schechter and  J. W.~F. Valle, \pr{D25}{82}{2951}

\bibitem{Avignone}
H. Klapdor, \ppnp{32}{94}{261}.

\bibitem{haxtongranada}
W. Haxton, \nps{31}{93}{88}.

\bibitem{QDN}
 L. Wolfenstein, \np{B186}{81}{147};
 J. W.~F. Valle, \pr{D27}{83}{1672} and references therein

\bibitem{KT}
E. Kolb, M. Turner, {\it The Early Universe},
Addison-Wesley, 1990, and references therein

\bibitem{ma1}
J. Bardeen, J. Bond and G. Efstathiou,\apj{321}{87}{28};
J. Bond and G. Efstathiou, \pl{B265}{91}{245}; 
M. Davis et al., \nat{356}{92}{489};
S. Dodelson, G. Gyuk and M. Turner, \prl{72}{94}{3754};
H. Kikuchi and E. Ma, \pr{D51}{95}{296}; 
H. B. Kim and J. E. Kim, \np{B433}{95}{421};
M. White, G. Gelmini and J. Silk, \pr{D51}{95}{2669}
A. S. Joshipura and J. W. F. Valle, \np{B440}{95}{647}.

\bibitem{SF}
A.D. Dolgov, these proceedings.

\bibitem{CMP}
Y. Chikashige, R. Mohapatra, R. Peccei, \prl{45}{80}{1926}

\bibitem{fae}
For recent reviews see J. W. F. Valle, 
{\it Gauge Theories and the Physics of Neutrino Mass}, 
\ppnp{26}{91}{91-171} (ed. A. Faessler); and G. Gelmini 
and S. Roulet, UCLA/94/TEP/36 and references therein.

\bibitem{774}
J. Schechter, J. W. F. Valle, \pr {D25} {82} {774}

\bibitem{V} 
J. W. F. Valle, \pl {B131} {83}{87};
G. Gelmini, J. W. F. Valle, \pl {B142} {84}{181};
J. W. F. Valle, \pl {B159} {85}{49};
M. C. Gonzalez-Garcia, J. W. F. Valle, \pl {B216} {89} {360}.
A. Joshipura, S. Rindani, \pr{D46}{92}{3000}

\bibitem{RPMSW}
J. C. Rom\~ao and J. W. F. Valle.
\pl{B272}{91}{436}; \np{B381}{92}{87}.

\bibitem{KTCS91}
E. W. Kolb, M. S. Turner, A. Chakravorty and D. N. Schramm, \prl{67}{91}{533}.

\bibitem{DI93}
A.D. Dolgov and I.Z. Rothstein, \prl{71}{93}{476}.

\bibitem{unstable}
M. Kawasaki, G. Steigman and H.-S. Kang, \np{B402}{93}{323}, 
\np{B419}{94}{105}; S. Dodelson, G. Gyuk and M.S. Turner,
\pr{D49}{94}{5068}; for a review see G. Steigman; in
{\sl Cosmological Dark Matter}, (World Scientific, 1994), 
ed. A. Perez, and J. W. F. Valle, p. 55

\bibitem{DPRV}
A.D. Dolgov, S. Pastor, J.C. Rom\~ao and J.W.F. Valle, in
preparation.
 
\bibitem{MASIpot3}
A Masiero, J. W. F. Valle, \pl {B251}{90}{273};
J. C. Romao,  C. A. Santos, and  J. W. F. Valle, \pl{B288}{92}{311};
A. Joshipura and  J. W.~F. Valle, \np{B397}{93}{105}

\bibitem{Pastor}
S. Pastor, \tp

\bibitem{cobe}
G.~F. Smoot et~al., \apj{396}{92}{L1-L5};
E.L.~Wright et al., \apj{396}{92}{L13}

\bibitem{iras}
R. Rowan-Robinson, in {\sl Cosmological Dark Matter}, 
(World Scientific, 1994), ed. A. Perez, and J. W. F. Valle, p. 7-18

\bibitem{cobe2}
N. Vittorio, these proceedings.

\bibitem{Caldwell}
D. Caldwell, talk at ICHEP94, Glasgow, 1994, preprint UCSB-HEP-95-03

\bibitem{cl}
B.T. Cleveland {\it et al.}, \nps{38}{95}{47}. 

\bibitem{k}
Y. Suzuki, \nps{B38}{95}{54}

\bibitem{ga}
GALLEX Collaboration,  P. Anselmann {\it et al.}, 
LNGS Report 95/37 (June 1995). 

\bibitem{sa}
SAGE  Collaboration,  J.S. Nico {\it et al.}, 
{\em Proc. 27th Conf. on High Energy Physics}, Glasgow, UK (July 1994). 

\bibitem{SSM}
J. N. Bahcall and R. K. Ulrich, \rmp{60}{90}{297};
J. N. Bahcall and M. H. Pinsonneault, \rmp{64}{92}{885}; 
J. N. Bahcall and M. H. Pinsonneault, preprint IASSNS-AST 95/24

\bibitem{CF}
J. N. Bahcall, \pl{B338}{94}{276};
V. Castellani, {\it et al} \pl{B324}{94}{245};
N. Hata, S. Bludman, and P. Langacker, \pr{D49}{94}{3622};
V. Berezinsky, {\rm Comments on Nuclear and Particle Physics} {\bf 21} 
(1994) 249

\bibitem{MSW}
M. Mikheyev, A. Smirnov, \sjnp{42}{86}{913};
L. Wolfenstein, \pr {D17}{78}{2369};\ib{D20}{79}{2634}.

\bibitem{noise} 
H. Nunokawa, A. Rossi, V. Semikoz and J. W. F. Valle, hep-ph/9602307

\bibitem{BalantekinLoreti} 
F. N. Loreti and A. B. Balantekin, \pr{D50}{94}{4762}

\bibitem{rossi}
A. Rossi,  \tp

\bibitem{Barish}
For a summary of the situation see T. Stanev, these proceedings;
and B. Barish, in proceedings of Int. Workshop on Elementary
Particle Physics: Present and Future, ed. A. Ferrer and
J. W. F. Valle, World Scientific, in press.

\bibitem{atm}
A. Suzuki, invited talk at SUSY 95, preprint TOHOKU-HEP-95-03

\bibitem{GRS}
 M Gell-Mann, P Ramond, R. Slansky,
in {\sl Supergravity},  ed. D. Freedman et al. (1979);
 T. Yanagida,
 in {\sl KEK lectures},  ed.  O. Sawada et al. (1979)

\bibitem{LR}
R.N.~Mohapatra and G.~Senjanovic, \pr{D23}{81}{165}
and references therein.

\bibitem{caldwell}
D.O.~Caldwell and R.N.~Mohapatra, \pr{D48}{93}{3259};
A. S. Joshipura, \zp{C64}{94}{31};
S. T. Petcov, A. Smirnov, \pl{B322}{94}{109}

\bibitem{2227}
 J. Schechter and  J. W. F. Valle, \pr{D22}{80}{2227}

\bibitem{DEG}
A. Ioannissyan, J.W.F. Valle, \pl{B332}{94}{93-99};
B. Bamert, C.P. Burgess, \pl{B329}{94}{289};
D. Caldwell and R. N. Mohapatra, \pr{D50}{94}{3477};
D. G. Lee and R. N. Mohapatra, \pl{B329}{94}{463}; 
A. S. Joshipura, \pr{D51}{95}{1321}

\bibitem{OLD}
 J. Schechter and  J. W. F. Valle, \pr{D21}{80}{309}

\bibitem{Choudhury}
See, for example, D. Choudhury et al \pr{D50}{94}{3486}

\bibitem{DARK92}
J.~T. Peltoniemi, D.~Tommasini, and J~W~F Valle,
\pl {B298}{93}{383}

\bibitem{DARK92B}
J.~T. Peltoniemi, and J~W~F Valle, \np{B406}{93}{409}; 
E. Akhmedov, Z. Berezhiani, G. Senjanovic and Z. Tao, 
\pr{D47}{93}{3245}.

\bibitem{DARK92D}
J.~T. Peltoniemi, \mpl{A38}{93}{3593}

\bibitem{JV95}
A. S. Joshipura, J. W. F. Valle, \np{B440}{95}{647}.

\bibitem{SST}
R. Mohapatra, J. W. F. Valle, \pr {D34} {86} {1642};
J. W. F. Valle, \nps{B11} {89} {118-177}

\bibitem{CERN}
M. Dittmar, M. C. Gonzalez-Garcia, A. Santamaria, J. W. F. Valle, 
\np{B332}{90}{1};
M. C. Gonzalez-Garcia, A. Santamaria, J. W. F. Valle, \ib{B342} {90} {108}.

\bibitem{lacasta}
C. Lacasta, Doctoral Thesis, Valencia University, 1995;
J. Fuster et al, paper in preparation. See also OPAL collaboration, 
\pl{B247}{90}{448} and \pl{B295}{92}{371}

\bibitem{alephmono}
ALEPH collaboration, \pl{B334}{94}{244}

\bibitem{BER}
J. Bernabeu, A. Santamaria, J. Vidal, A. Mendez, J. W. F. Valle, 
\pl {B187} {87} {303}; J. G. Korner, A. Pilaftsis, K. Schilcher,
\pl {B300} {93} {381}

\bibitem{CP}
G. C. Branco, M. N. Rebelo, J. W. F. Valle, \pl {B225} {89} {385}; 
N. Rius, J. W. F. Valle, \pl{B246}{90}{249}

\bibitem{3E}
M. C. Gonzalez-Garcia, J. W. F. Valle, \mpl{A7}{92}{477};
erratum \mpl{A9}{94}{2569}; A. Ilakovac, A. Pilaftsis, 
\np{B437}{95}{491}; A. Pilaftsis, \mpl{A9}{94}{3595}

\bibitem{TTTAU}
R. Alemany et. al. hep-ph/9307252, published in ECFA/93-151, 
ed. R. Aleksan, A. Ali, p. 191-211 

\bibitem{ETAU}
M. Dittmar, J. W. F. Valle,
contribution to the High Luminosity at LEP working group, 
yellow report CERN-91/02, p. 98-103 

\bibitem{opallfv}
Opal collaboration, \pl{B254}{91}{293} and \zp{C67}{95}{365}; 
L3 collaboration, \prep{236}{93}{1-146}; \pl{B316}{93}{427}, 
Delphi collaboration, \pl{B359}{95}{411}.

\bibitem{zee.Babu88}
A. Zee, \pl{B93}{80}{389}; K.~S. Babu, \pl{B203}{88}{132} 

\bibitem{JoshipuraValle92}
 A. Joshipura and  J. W.~F. Valle, \np{B397}{93}{105}
and references therein
 J.~C. Romao,  F. de~Campos, and  J. W.~F. Valle,
\pl{B292}{92}{329}.
A. S. Joshipura, S. Rindani, \prl{69}{92}{3269};
 R. Barbieri, and L. Hall, \np{B364}{91}{27};
G. Jungman and M. Luty, \np{B361}{91}{24};
E. D. Carlson and L. B. Hall, \pr{D40}{89}{3187};
S. Bertolini, A. Santamaria, \pl{B213}{88}{487}

\bibitem{alfonso}
A. Lopez-Fernandez, J. Romao, F. de Campos and  J. W.~F. Valle,
\pl{B312}{93}{240};
B. Brahmachari, A. Joshipura, S. Rindani, D. P. Roy, K. Sridhar,
\pr{D48}{93}{4224};
F. de Campos et al., proceedings of Moriond94, hep-ph/9405382

\bibitem{ebolepp2}
F. de~Campos, O. Eboli, J. Rosiek, J. W. F. Valle, 
hep-ph/9601269

\bibitem{HA}
F. de~Campos, M. A. Garcia-Jare\~no, A. Joshipura, J. Rosiek, 
J. W. F. Valle, and D. P. Roy, \pl{B336}{94}{446}

\bibitem{EE500}
O. Eboli, M. C. Gonzalez-Garcia, A. Lopez-Fernandez, S. F. Novaes,
J. W. F. Valle, \np{B421}{94}{65}
 
\bibitem{granada}
D. Choudhhury, and D. P. Roy, \pl{B322}{94}{368};
J.~C. Romao,   F. de~Campos, L. Diaz-Cruz,  and  J. W.~F. Valle,
\mpl{A9}{94}{817}; J. Gunion, \prl{72}{94}{199}

\end{thebibliography}

\end{document}